\documentclass[11pt,a4paper]{article}

\usepackage{geometry}
\usepackage[applemac]{ inputenc}

\usepackage{mathrsfs}
\usepackage{amsmath}
\usepackage{amssymb}
\usepackage{multicol}
\usepackage{float}
\usepackage{makeidx}
\usepackage{layout}
\usepackage{array}
\usepackage{boxedminipage}
\usepackage{hyperref}
\usepackage{latexsym}
\usepackage{indent first}

\usepackage{subcaption,graphicx}
\usepackage{appendix}

\usepackage{color}

\usepackage[sectionbib,square]{natbib}

\usepackage[normalem]{ulem}

\DeclareMathOperator{\sech}{sech}

\title{Solitary water wave interactions for the Forced Korteweg-de Vries equation}
\author{M. V. Flamarion$^{1}$ and R. Ribeiro-Jr$^{2}$}

\date{}

\begin{document}

\maketitle

{\footnotesize
	\begin{center}
	$^{1}$ UFRPE/Rural Federal University of Pernambuco, UACSA/Unidade Acad{\^e}mica do Cabo de Santo Agostinho, BR 101 Sul, 5225, 54503-900, Ponte dos Carvalhos, Cabo de Santo Agostinho, Pernambuco, Brazil.
	
		$^2$ UFPR/Federal University of Paran\'a,  Departamento de Matem\'atica, Centro Polit\'ecnico, Jardim das Am\'ericas, Caixa Postal 19081, Curitiba, PR, 81531-980, Brazil. 
	
	\end{center}
	
	}
\begin{abstract}
The aim of this work is to study solitary water wave interactions between two topographic obstacles for the forced Korteweg-de Vries equation (fKdV). Focusing on the details of the interactions,   we identify regimes in which solitary wave interactions maintain two well separated crests and regimes where the  number of local maxima varies according to the laws $2\rightarrow 1\rightarrow 2\rightarrow 1\rightarrow 2$  or $2\rightarrow 1\rightarrow 2$. It shows that the geometric Lax-categorization of Korteweg-de Vries (KdV) two-soliton interactions still holds for the fKdV equation.

\vspace{0.25 cm}

 
\end{abstract}

\section{Introduction}
The forced Korteweg-de Vries equation (fKdV) has been used as a model to describe atmospheric flows encountering topographic obstacles, flow of water over rocks (\cite{Baines}),  ship waves and  ocean waves generated by storms (when a low pressure region moves on the surface of the ocean \cite{Johnson}). 

Solitary waves have a wide range of applications, for instance, in water waves, optical fibers, superconductive electronics, elementary-particle physics, quantum physics and more recent applications in biology and cosmology (\cite{Joseph}).  It is well known that the Korteweg-de Vries equation (KdV) is used to describe the propagation and interaction between solitary waves. Studying numerical solutions of the KdV equation \cite{Zabusky} were the first to observe that solitary waves interact during the collision and return to its initial form. They named this type of waves as solitons. This study raised interest to investigate further details of soliton interactions. Since then many works have been done on this topic. It is hard to give a comprehensive overview of contributions. For the interested reader, we mention a few articles which are seminal in this field.

 \cite{Lax}  classified overtaking collisions of two solitons  in three categories according to the number of crests observed during the interaction. More precisely, he proved that the type of the collision can be classified according to the ratio of the initial amplitude of the solitons. 
 The categorization given by Lax was verified experimentally by \cite{Maxworthy} and numerically by \cite{Su}  for a higher order model. More recently,  \cite{Craig} presented a work in which is given a broad review on solitary wave interactions. They investigated numerically and experimentally solitary wave collisions for the Euler equations. Their numerical simulations show that the collisions of two solitary waves fit into the three 
geometric categories of the KdV two-soliton solutions defined by Lax. However, the algebraic classification based on the ratio of the initial amplitudes is within a different range of the one considered by Lax.

In this paper we  investigate numerically in details the interaction of two solitary wave solutions of the fKdV. More precisely,  we analyse the interaction of these two waves between obstacles. We find the three geometric categories described by \cite{Lax} for the KdV two-soliton interaction. However, our experiments  indicate that an algebraic categorization similar to the one presented by Lax is not possible for the fKdV.

This article is organized as follows. In section 2 we present the mathematical formulation of the non-dimensional fKdV equation.  The results are presented in section 3 and the conclusion in section 4.

\section{The forced Korteweg-de Vries equation}
We consider an inviscid, incompressible, homogeneous fluid on a shallow channel with variable topography in the presence of a constant current. The flow of the fluid can be classified by the Froude number ($F$), which is defined by the ratio of the upstream velocity and the critical speed of shallow water. When the Froude number is near critical ($F\approx 1$), and the amplitude of the  topography is small the weakly nonlinear, weakly dispersive model given by the dimensionless forced Korteweg-de Vries equation
\begin{equation}\label{fKdV}
\zeta_{t}+f\zeta_{x}-\frac{3}{2}\zeta\zeta_{x}-\frac{1}{6}\zeta_{xxx}=\frac{1}{2} h_{x}(x),
\end{equation}
is used to describe the flow over the obstacle (\cite{Pratt, Wu1, Grimshaw, Paul, Marcelo-Paul-Andre}). Here, $\zeta(x,t)$ is the free-surface displacement over the undisturbed surface and $h(x)$ is the obstacle submerged. The parameter $f$ represents a perturbation of the Froude number, i.e, $F=1+\epsilon f$, where $\epsilon>0$ is a small parameter.  

It is important to point out that the equation (\ref{fKdV}) conserves mass $(M(t))$, with
$$\frac{dM}{dt}=0,  \mbox{   \;\;\; where  \;\;\;\ }  M(t) = \int_{-\infty}^{+\infty}\zeta(x,t)\, dx.$$

When the bottom is flat ($h_x=0$) a traveling solitary wave solution for (\ref{fKdV}) is 
\begin{equation}\label{solitary}
\zeta(x,t)=A\sech^{2}(k(x-ct)),  \;\ A=\frac{4}{3}k^{2}, \;\ c =f-\frac{1}{2}A.
\end{equation}
Notice that when $f=A/2$ the solution is stationary.

The fKdV equation (\ref{fKdV}) is solved numerically using a Fourier pseudospectral method with an integrating factor for the linear part, thus avoiding numerical problems due to the higher-order dispersive term. We consider the computational domain  with an uniform grid. All derivatives in $x$ are computed spectrally (\cite{Trefethen}).  Besides, the time evolution is calculated through the Runge-Kutta fourth-order method (\cite{Marcelo-Paul-Andre}).

\section{Results}

We investigate collisions of two solitary waves between the osbtacles. For this purpose the initial condition of (\ref{fKdV}) is given by a linear sum of two well-separate solitary waves $$\zeta(x,0)=S_{1}\sech^{2}(k_{1}(x-\phi))+S_{2}\sech^{2}(k_{2}(x+\psi)),$$ 
where $S_{1}=4k_{1}^{2}/3$,  $S_{2}=4k_{2}^{2}/3$, and $\phi,\psi$ are positive constants.
Our focus is to categorize the  collision of two solitary waves  into three types in the same spirit as presented in \cite{Lax} for the KdV equation. Studying overtaking ($S_{2}>S_{1}$) collisions Lax has classified the details of two-soliton interactions as follows:
\begin{itemize}
	\item[{\bf (A)}] For any time $t$ the solution of the KdV has two well-defined and separate crests, and it happens when $S_{2}/S_{1}<(3+\sqrt{5})/2\approx 2.62$.
		\item[{\bf (B)}]  		During the collision the number of local maxima varies according to $2\rightarrow 1\rightarrow 2\rightarrow 1\rightarrow 2$, and for this case we have $(3+\sqrt{5})/2<S_{2}/S_{1}<3$.
	\item[{\bf(C)}]  In the interaction the number of local maxima changes as $2\rightarrow 1\rightarrow 2$ and it occurs for $S_{2}/S_{1}>3$.
\end{itemize}
After the collision the main notable feature is that the waves are phase shifted, i.e., their crest are slightly shifted from the trajectories of the incoming centers.

 In the following simulations we use the parameters $\beta=20$, $\epsilon=0.01$ and $f=0.34$. Besides, in order to avoid radiation from the topography we sum a term $r(x)$ to the initial condition, where $r$ is the steady solution of the uniform flow.

\begin{figure}[h!]
	\centering	
	\includegraphics[scale =1]{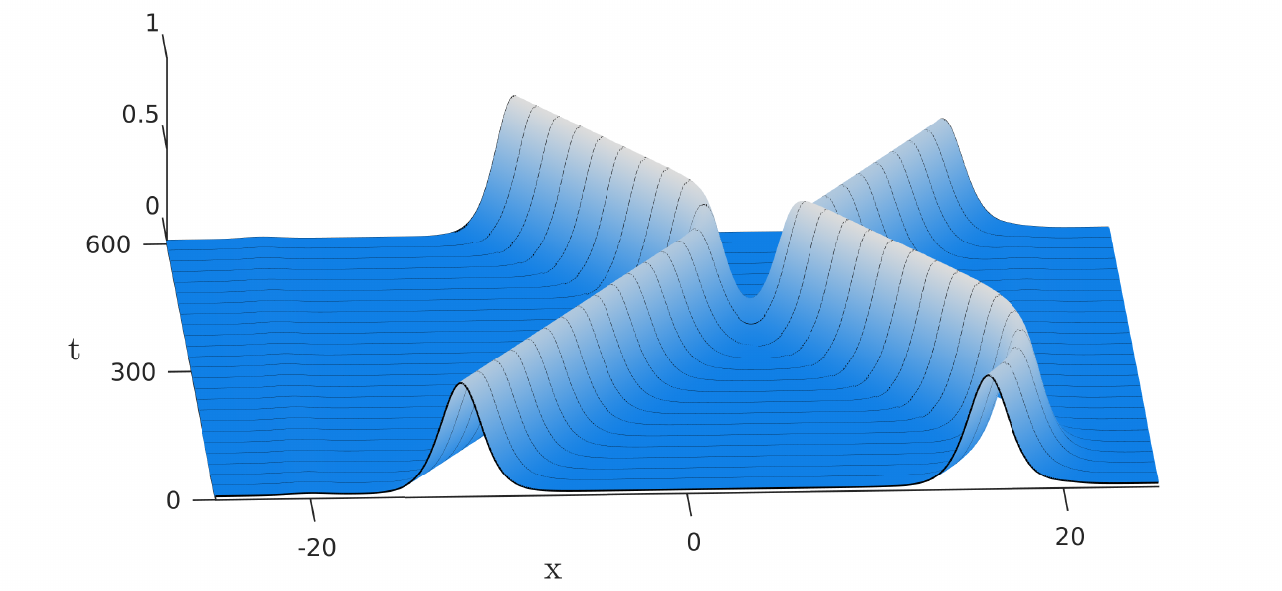}
	\includegraphics[scale =1]{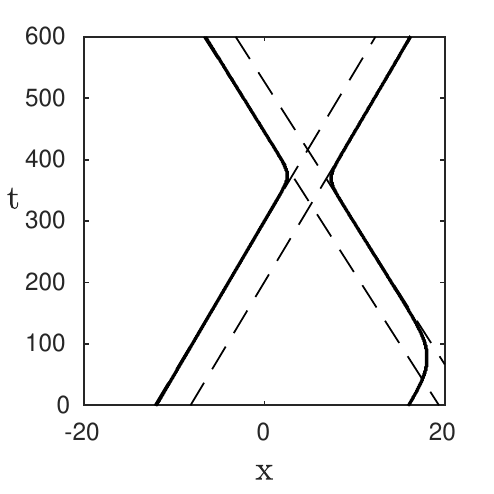}
	\includegraphics[scale =1]{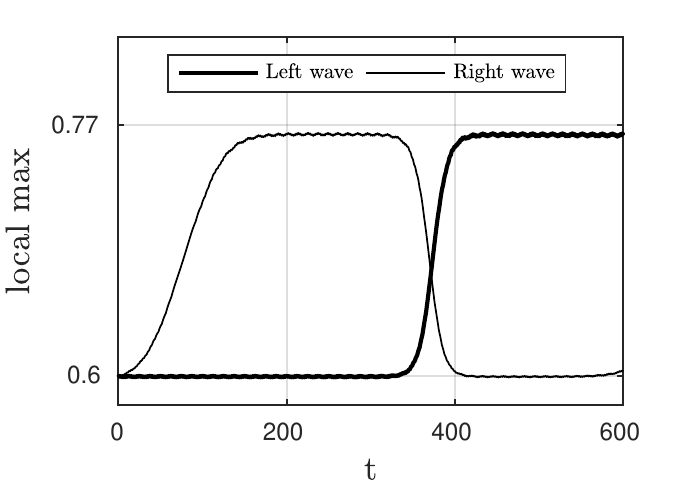}
	\caption{Top: Collision of two well-separate solitary waves -- category {\bf(A)}. Bottom (left): Crest trajectories. Bottom (right): The local maxima of the solution as a function of time. Parameters $S_{1}=S_{2}=0.6$, $\phi=16$ and $\psi=12$.}
	\label{colisaoA}
\end{figure}
We start considering the collision of two well-separeted solitary waves that initially have the same amplitude. Details of the wave profile are given in figure \ref{colisaoA} (top).
Initiallly two solitary waves propagate downstream. When the right wave reaches the obstacle its amplitude increases and the wave reflects back upstream. Then the waves collide mimicking a counterpropagating collision. As the right wave approaches the wave with smaller crest the larger wave begins to shrink and the smaller one begins to grow until the two waves interchange their roles (see figure \ref{colisaoA} (bottom-right). Throughout the interaction there are two well-defined and separate crests as shown in figure \ref{colisaoA} (bottom-left). This behaviour is similar to  case {\bf (A)} of Lax classification. Figure \ref{colisaoA2} displays the continuation of figure \ref{colisaoA}(top). After a series of collisions both waves escape out. We point out that the numerical method conserves mass and the relative error is:
$$\frac{\displaystyle\max_{0\le t\le 10^{4}}|M(t)-M(0)|}{|M(0)|}=\mathcal{O}(10^{-16}).$$

\begin{figure}[h!]
	\centering	
	\includegraphics[scale =0.99]{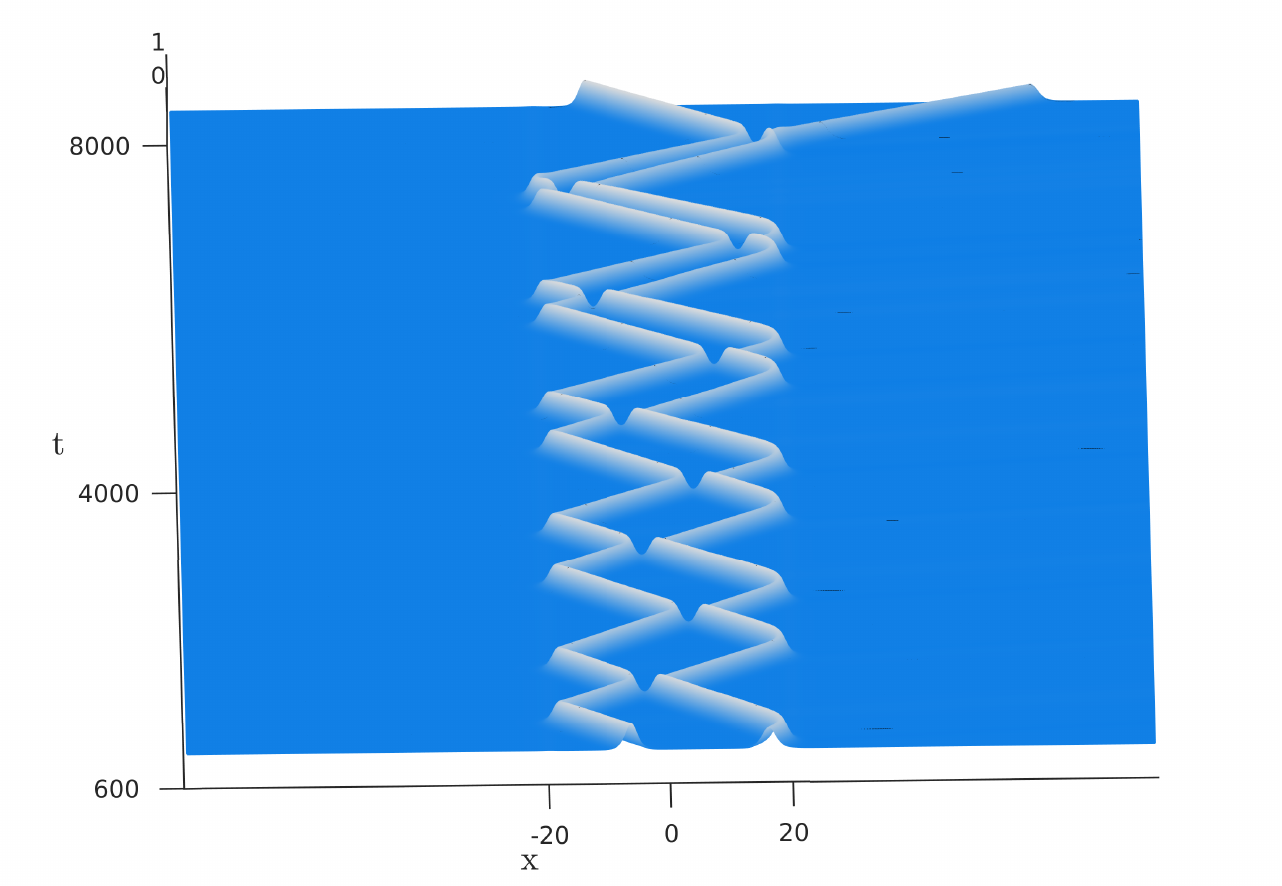}
	\caption{Continuation of figure \ref{colisaoA} (top). Parameters $S_{1}=S_{2}=0.6$, $\phi=16$ and $\psi=12$.}
	\label{colisaoA2}
\end{figure}

\begin{figure}[h!]
	\centering	
	\includegraphics[scale =1]{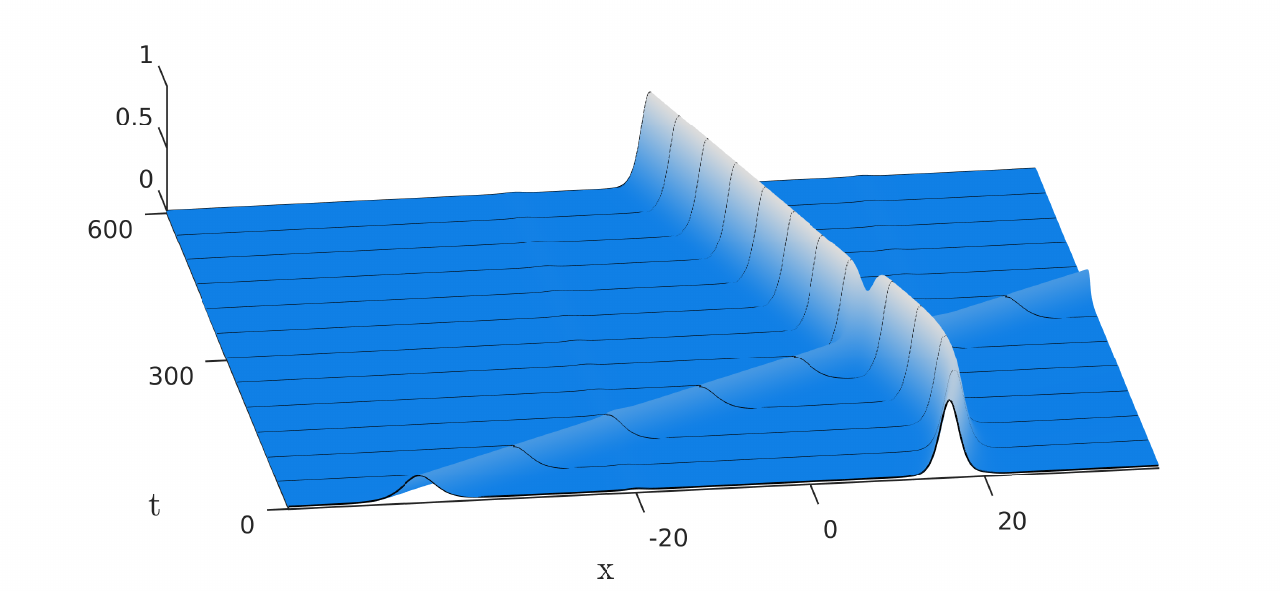}
	\includegraphics[scale =1]{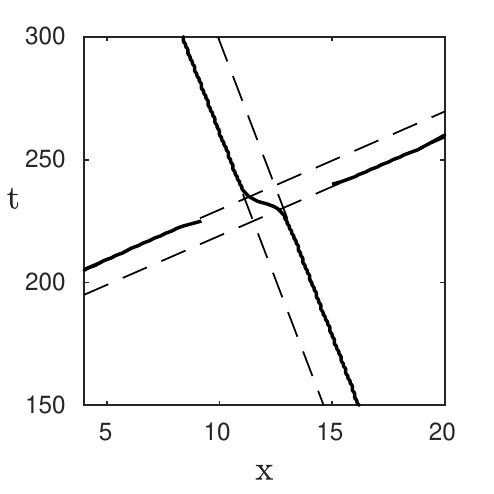}
	\includegraphics[scale =1]{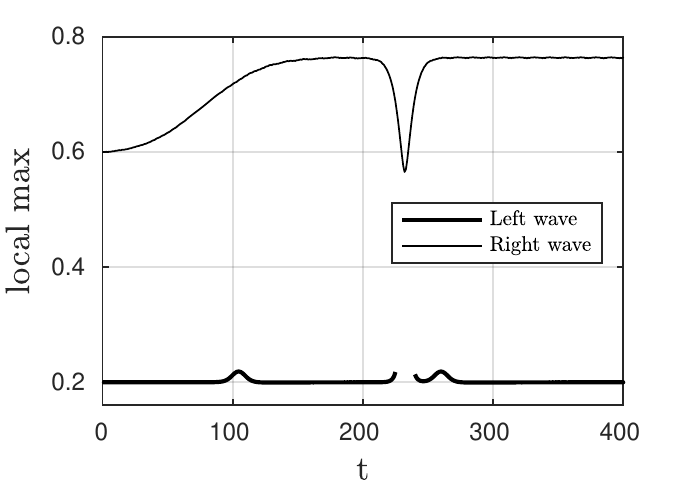}
	\caption{Top: Collision of two well-separate solitary waves -- category {\bf(C)}. Bottom (left): Crest trajectory. Bottom (right): The local maxima of the solution as a function of time. Parameters $S_{1}=0.6$, $S_{2}=0.2$, $\phi=16$ and $\psi=45$.}
	\label{colisaoC}
\end{figure}

Figure \ref{colisaoC} (top) displays the collision of two well-separate solitary waves that initially have different amplitudes.  Differently from the previous case there is a period of time in the interaction that only one crest exists. The interaction is characterized by an absorption of the smaller wave and its reemission later, along with a phase lag in the trajectories of the crest, see figure \ref{colisaoC} (bottom).

\begin{figure}[h!]
	\centering	
	\includegraphics[scale =1]{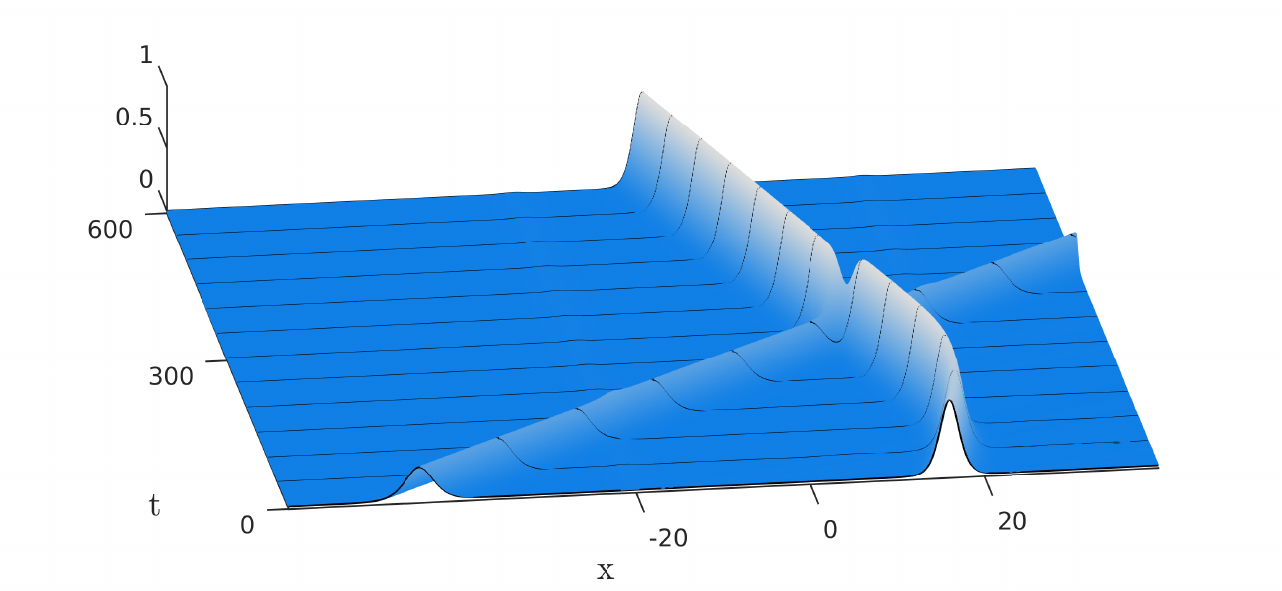}
	\includegraphics[scale =1]{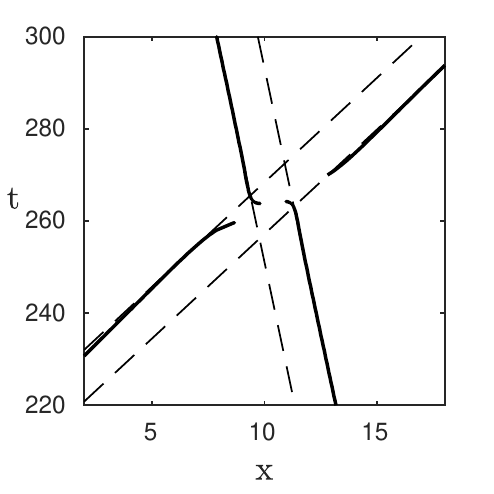}
	\includegraphics[scale =1]{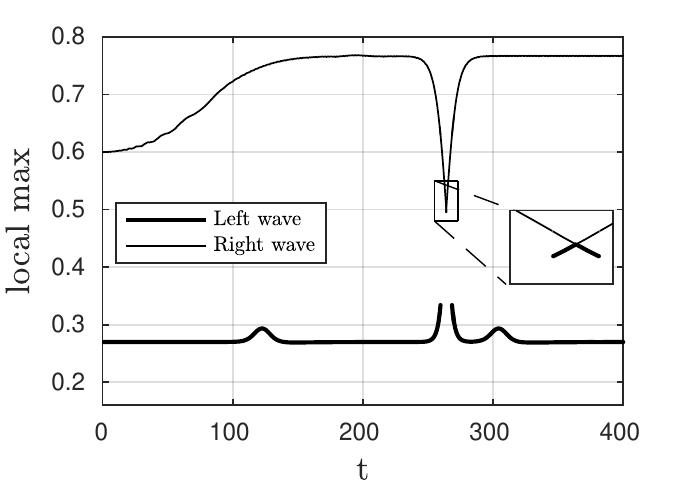}
	\caption{Top: Collision of two well-separate solitary waves -- category {\bf(B)}. Bottom (left): Crest trajectory. Bottom (right): The local maxima of the solution as a function of time. Parameters $S_{1}=0.6$, $S_{2}=0.27$, $\phi=16$ and $\psi=45$.}
	\label{colisaoB}
\end{figure}

\begin{figure}[h!]
	\centering
	\includegraphics[scale =1]{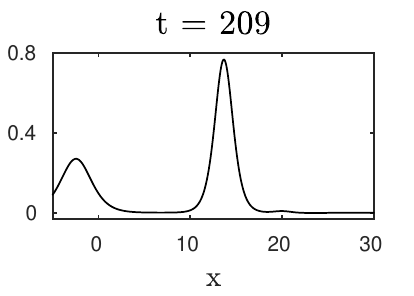}	
	\includegraphics[scale =1]{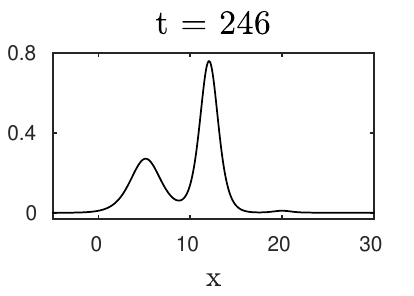}
	\includegraphics[scale =1]{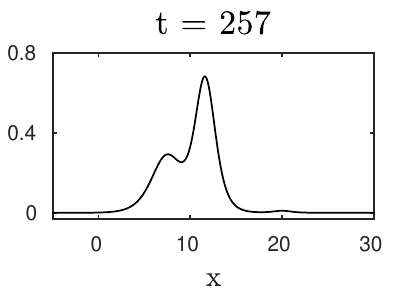}
	\includegraphics[scale =1]{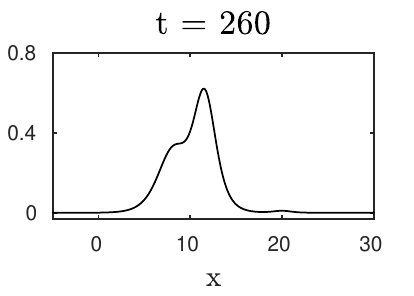}
	\includegraphics[scale =1]{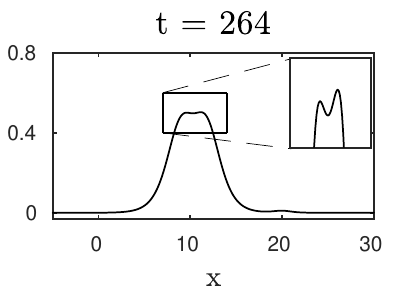}
	\includegraphics[scale =1]{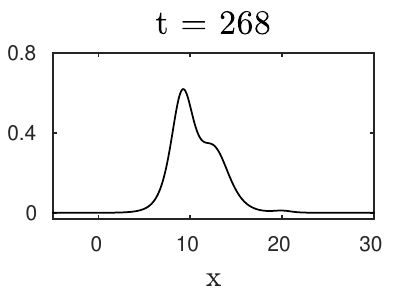}
	\includegraphics[scale =1]{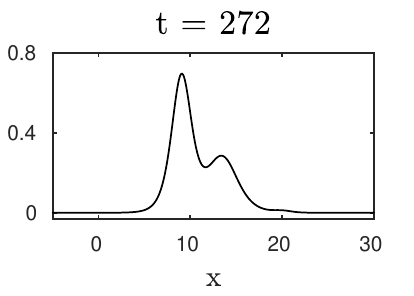}
	\includegraphics[scale =1]{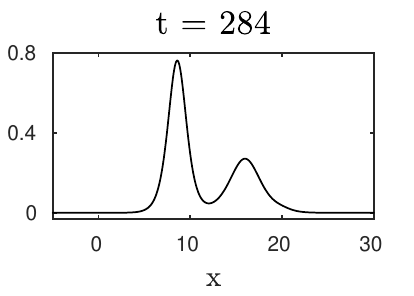}
	\includegraphics[scale =1]{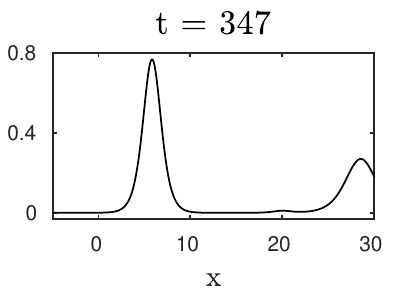}
	\caption{Snapshots of the interaction of the two well-separate solitary waves of figure \ref{colisaoB} during the collision -- category {\bf(B)}. }
	\label{colisaoBBB}
\end{figure}

Lastly, we show a collision that presents features similar to the cases {\bf (A)} and {\bf (B)} simultaneously, see figure \ref{colisaoB}.  The smaller wave is first swallowed, then expelled by the larger one. This dynamic is very similar to the description given previously in case {\bf (C)}. However, during the collision there is a central region consisting of two crests. This behaviour is described in great details in a serie of snapshots depicted in  figure \ref{colisaoBBB}.    

For the KdV equation the transition between two categories is determined by the ratio of the amplitudes of the two separated solitary waves given initially. 
However, for the fKdV is not possible to estimate a similar condition regarding the ratio of the amplitudes as shown in table \ref{table1}. Nevertheless  the fKdV equations still holds the geometric features of the Lax categorization.
\begin{table}[h!]
\centering
\begin{tabular}{c|c|c|c}\hline\hline
$S_{1}$ & $S_{2}$ & $\max{\{S_{1},S_{2}}\}/\min\{{S_{1},S_{2}}\}$ & category \\   \hline
0.60  &	0.30 &	 2.00	& \bf{A} \\ \hline
0.60	& 0.29 &	2.06 &	\bf{A} \\ \hline
0.60	& 0.28 &	2.14 & \bf{B} \\ \hline 
0.60	& 0.27 &	2.22	& \bf{B} \\ \hline 
0.60	& 0.26 &	2.30 &	\bf{B} \\ \hline
0.60	& 0.25	& 2.40	& \bf{B}  \\ \hline
0.60 & 	0.24 &	2.50 & 	\bf{C} \\ \hline 
0.60 & 	0.20 & 	3.00 &	\bf{C} \\  \hline
0.55	& 0.26	& 2.06&	\bf{B} \\ \hline
0.55 &	0.27 &	2.00 &	\bf{B} \\  \hline
0.55	& 0.37 &1.50&	\bf{A} \\ \hline
\end{tabular}
\caption{Classification of the collision for different values of $S_1$ and $S_2$.}\label{table1}
\end{table}

\section{Conclusion}
In this paper we have investigated solitary wave  collisions for the fKdV equation. Through a pseudospectral numerical method, we showed that the geometric Lax characterisation for the KdV two-soliton interaction still holds for the fKdV, i.e. solitary wave interactions maintain two well separated crests in regime {\bf (A)}, the larger solitary wave absorbs the smaller one and the  number of local maxima varies according to the law $2\rightarrow 1\rightarrow 2\rightarrow 1\rightarrow 2$ in regime {\bf (B)} or  the number of local maxima changes as $2\rightarrow 1\rightarrow 2$, case {\bf (C)}. Although there  are a number of theoretical and numerical works on collisions for the KdV equation, as far as we know there are no articles focused  on collision details for the fKdV equation.

\section{Acknowledgements}

The authors are grateful to IMPA-National Institute of Pure and Applied Mathematics for the research support provided during the Summer Program of 2020 to Prof. Paul Mileswki (University of Bath)  for his constructive comments and suggestions which improved the manuscript. M.F. is grateful to Federal University of Paran{\' a} for the visit to the Department of Mathematics. R.R.-Jr is grateful to University of Bath for the extended visit to the Department of Mathematical Sciences.

\bibliographystyle{abbrv}

\end{document}